\documentclass[aps,prl,twocolumn,floats,showpacs]{revtex4}
\usepackage{graphicx}
\usepackage{amsmath}
\usepackage{amssymb}
\usepackage{color}

\newcommand{\pipulse}{{$\pi$-pulse }}

\begin{document}
\title{Strong-field effects in Rabi oscillations between a single state and a superposition of states}

\date{\today}
\author{S.~Zhdanovich$^{1,3}$, J.W.~Hepburn$^{1,2,3}$, and V.~ Milner$^{1,3}$}
\affiliation{Departments of  Physics \& Astronomy$^1$ and Chemistry$^2$, and The Laboratory for Advanced Spectroscopy and Imaging Research (LASIR)$^3$, The University of British Columbia, Vancouver, Canada}

\begin{abstract}{Rabi oscillations of quantum population are known to occur in two-level systems driven by spectrally narrow laser fields. In this work we study Rabi oscillations induced by shaped broadband femtosecond laser pulses. Due to the broad spectral width of the driving field, the oscillations are initiated between a ground state and a coherent superposition of excited states, or a ``wavepacket'', rather than a single excited state. Our experiments reveal an intricate dependence of the wavepacket phase on the intensity of laser field. We confirm numerically that the effect is associated with the strong-field nature of the interaction, and provide a qualitative picture by invoking a simple theoretical model.}
\end{abstract}


\maketitle

Efficient transfer of population from one energy eigenstate of a quantum system to another eigenstate is an important tool in many fields of physics and chemistry such as control of molecular dynamics and chemical reactions \cite{ShapiroBrumerBook}, quantum computing and information processing \cite{NielsenChuangBook}, precision spectroscopy \cite{Dhar94}, cold and ultra-cold chemistry \cite{Doyle04} and nanoscience \cite{Kral07}. In the simplest case of a two-level atom interacting with a resonant electro-magnetic field, the dynamics of the atomic population, described by Maxwell-Bloch equations, exhibit well known periodic Rabi oscillations \cite{AllenEberlyBook, Shore1990}. The phase of these oscillations is directly related to the notion of ``pulse area''. When the latter assumes the value of $\pi$ (so-called ``\pipulse''), the transfer of quantum population between the two levels is complete.

Rabi oscillations serve as a convenient calibration tool for measuring the pulse area and excited state population. Being a function of the pulse duration, intensity, detuning from resonance, and the transition dipole moment, an experimentally measured pulse area enables retrieving one of these parameters if the others are known. The contrast of Rabi oscillations reflects the degree of coherence of an atom-photon interaction, and can be utilized for assessing the coherence properties of either the atomic (molecular) system or the applied electromagnetic field.

Although $\pi$-pulses and Rabi oscillations between metastable states of atoms and molecules are observed and exploited quite routinely, e.g. between Rydberg states \cite{Sirko1990, Rempe1987} or spin states in Bose-Einstein condensates \cite{Wright09}, population oscillations between two electronic states are much harder to detect due to quick, typically nanosecond, spontaneous decay of electronic coherence \cite{Reetz08}. Shortening the excitation pulses beyond this time scale results in two complications. First, nonlinear effects, such as AC Stark shift and multi-photon ionization, become non-negligible as their efficiency increases with decreasing length of a $\pi$-pulse \cite{Trallero-Herrero06}. Second, the spectral width of short pulses becomes comparable with the energy level spacing in the atomic or molecular spectrum, making the two-level approximation invalid. In the regime when the population of the quantum states participating in the interaction changes substantially, the perturbative approach, often invoked in ultrafast quantum electronics \cite{Dudovich01, Prakelt2004}, is not applicable  \cite{Chuntonov2008, Trallero2007}. Hence, multiple excited states cannot be treated as independently driven by separate resonant frequency components of an excitation pulse.

Coherent population transfer with broadband laser pulses has been the focus of much experimental and theoretical work in the last two decades. The population dynamics and the origin of population oscillations are qualitatively different in weak and strong laser fields. In the case of an excitation by weak laser pulses (i.e. with pulse areas much smaller than $\pi $) driving a single photon transition, the final quantum state of an atomic system is defined solely by the resonant spectral component of the applied laser field, and the population of the target excited state is linearly proportional to the spectral power density at the transition frequency \cite{Dudovich05}. In this perturbative regime of interaction, oscillatory dynamics of the excited state population have been observed and attributed to coherent transients \cite{Zamith2001}. The latter have been used for quantum state reconstruction \cite{Monmayrant2006} and for enhancing the excited state population by means of femtosecond pulse shaping \cite{Dudovich2002}.

Two-photon weak-field transitions offer another mechanism of population oscillations due to the presence of resonant intermediate states. Quantum interferences arising from the evolution of the intermediate wavepacket have been studied and used for a temporal control of atomic population \cite{Blanchet1997}. Weak shaped femtosecond pulses have also been utilized to control two-photon transitions without a resonant intermediate state \cite{Meshulach98, Meshulach99, Prakelt2004}. Similarly to weakly driven single-photon transitions, the final excited state population in the case of a two-photon resonance is described by the second order perturbation theory, and is proportional to the resonant spectral component of the second harmonic. The latter depends on the phase shaping applied at the fundamental frequency, which may therefore lead to the population oscillations of the target state.

In this work we discuss the oscillations of atomic population in strong fields, governed by the non-perturbative regime of atom-photon interaction. Two-level atoms in strong laser fields have been thoroughly studied in a series of works on selective population of dressed states (SPODS \cite{Wollenhaupt2003, Wollenhaupt2006, Wollenhaupt2006A}) both with transform-limited and shaped femtosecond pulses. An oscillatory behavior of the excited state population as a function of the laser intensity has been observed in atomic Potassium \cite{Wollenhaupt2003} and Rubidium \cite{Zhdanovich08} providing direct evidence of femtosecond Rabi oscillations in a two-level system. Frequency chirping has been used to selectively populate dressed states of an atom subject to a strong laser field \cite{Wollenhaupt2006, Bayer2008}. Pulse trains have also been utilized for controlling population transfer in the strong-field interaction regime. Relative phase between the pulses in the train has been shown to control the adiabaticity of the population transfer process \cite{Zhdanovich08, Torosov11} and to provide selectivity in populating a single dressed state \cite{Wollenhaupt2006A, Wollenhaupt2010}.

When applied to multi-level systems, strong laser fields often shift the energy of near-resonant atomic levels, creating a number of time-dependent dressed states evolving on a femtosecond time scale \cite{Konorov2011}. When the energy of two such states become equal, the quantum system undergoes an avoided crossing. Quantum interferences between adiabatic and non-adiabatic routes through the avoided crossing result in the oscillations of the target state population \cite{Broers1992, Balling1994, Jones95, Maas1999, Krug2009}. In this strong-field interaction regime, frequency chirping and adaptive feedback loops have been exploited for increasing the population of a single target state  \cite{Trallero-Herrero06} up to a complete population transfer \cite{Krug2009}.

The ability to cover several target states simultaneously by the broad spectrum of an ultrashort laser pulse offers an opportunity to transfer population into a coherent superposition of states, or a ``wavepacket'', rather than a single target state. The oscillatory dynamics of the electronic wavepackets created by weak ultrashort pulses have been observed in a number of femtosecond pump-probe experiments \cite{Christian1993, Nicole1999, Zamith2000}. In these studies, the perturbative regime of interaction ensures that the wavepacket dynamics are independent on the strength of the applied laser field. Control of the wavepacket phase by means of a spectral phase shaping of pump pulses has been demonstrated \cite{Chatel2003, Chatel2008}.

In a series of recent works \cite{Shapiro09, Zhdanovich09}, we have demonstrated how a complete population transfer can be executed between a single state and a wavepacket. The method is based on a quasi-adiabatic evolution of the system in strong laser fields, as confirmed by the demonstrated robustness of the population transfer against uncertainty in laser intensity and wavelength. Surprisingly, despite the strong applied fields needed for the adiabatic passage, the phase of the excited wavepacket has been shown to obey the rules dictated by the perturbation theory. That is, similar to \cite{Chatel2003, Chatel2008}, changing the phase of the resonant spectral components of the excitation pulse resulted in the corresponding change of the wavepacket phase \cite{Zhdanovich09}. The effect stems from the following argument. For the adiabatic passage to work in the case of a multi-level target state, it has to be carried out in a piecewise manner with a train of mutually coherent femtosecond pulses. Even though the accumulative pulse area of the whole train is high ($>4\pi $), individual pulses in the train are relatively weak, which enables the perturbative control scheme of the target wavepacket.

The situation is qualitatively different if the wavepacket is created by a single strong laser pulse. Dynamic Stark shifts of the atomic levels forming the wavepacket result in the intensity dependent phase, which has to be taken into account on any route to strong-field coherent control \cite{Trallero2007}. In this paper, we present an experimental observation of Rabi oscillations between a single ground state and a spin-orbit electronic wavepacket, initiated by a strong unshaped laser pulse. We study the dependence of the wavepacket phase on the applied pulse area, and discuss the observed non-trivial step-like phase behavior using a simple model.


\begin{figure}
\centering
\includegraphics[width=0.98\columnwidth]{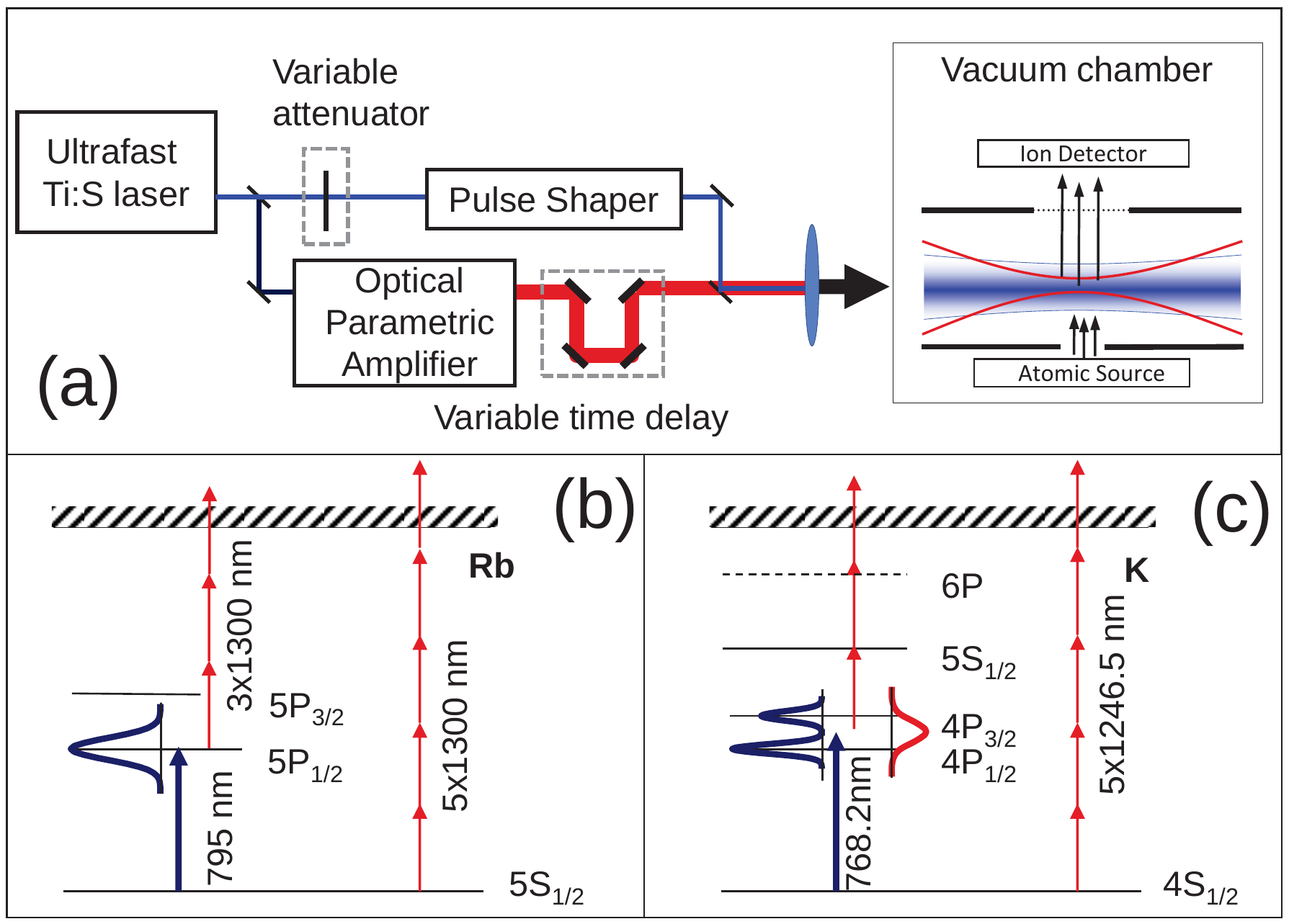}
\caption{(Color online) (a): Experimental setup (see text for details); (b) and (c): relevant energy levels and excitation paths for Rb and K, respectively. Thick blue and thin red lines denote excitation and probe photons, respectively.}
  \vskip -.1truein
  \label{FigSetup}
\end{figure}
Our experimental setup, shown in Fig.\ref{FigSetup}, has been described in our earlier paper \cite{Zhdanovich08} (see Fig.\ref{FigSetup}(a)). Briefly, it consists of a regenerative femtosecond Titanium-Sapphire amplifier producing 2mJ, 130 fs pulses at 1 KHz repetition rate and central wavelength around 800 nm. The laser beam is split into two parts. The first (``excitation'') beam is used to drive Rabi oscillations of atomic population between the ground and excited electronic state(s) (thick blue arrows in Fig.\ref{FigSetup}(b,c)). Excitation pulses are spectrally shaped with a home made liquid crystal based pulse shaper \cite{Weiner00}, and weakly focused onto a vapor cloud of either Rubidium or Potassium atoms inside a vacuum chamber (shaded blue beam in Fig.\ref{FigSetup}(a)). To determine the excited state population, we ionize the atoms with a weak 120 fs infrared ``probe'' pulse, generated by an optical parametric amplifier (OPA) pumped with the second part of the 800 nm beam. Probe pulses follow excitation pulses with a variable time delay, and are focused much tighter on the central part of the interaction region (unshaded red beam in Fig.\ref{FigSetup}(a)). We are able to detect the excited state population with good selectivity because the ionization of the ground state requires two more photons and is therefore negligibly weak (Fig.\ref{FigSetup}(b,c)).

In the case of a resonant excitation by a laser pulse with the electric field envelope $\varepsilon(t)$, the final population of the excited state oscillates as $\sin^2\left( A/2 \right)$ \cite{AllenEberlyBook}, where $A := \int_{-\infty}^{+\infty} dt |\Omega_0(t)|$ is the pulse area, $\Omega_0(t) := \varepsilon(t) \mu/\hbar$ is the time dependent Rabi frequency, and $\mu$ is the transition dipole moment. Since $A$ scales linearly with the field amplitude, the excited state population is expected to oscillate with pulse energy for a given pulse duration.

\begin{figure}
\centering
\includegraphics[width=0.98\columnwidth]{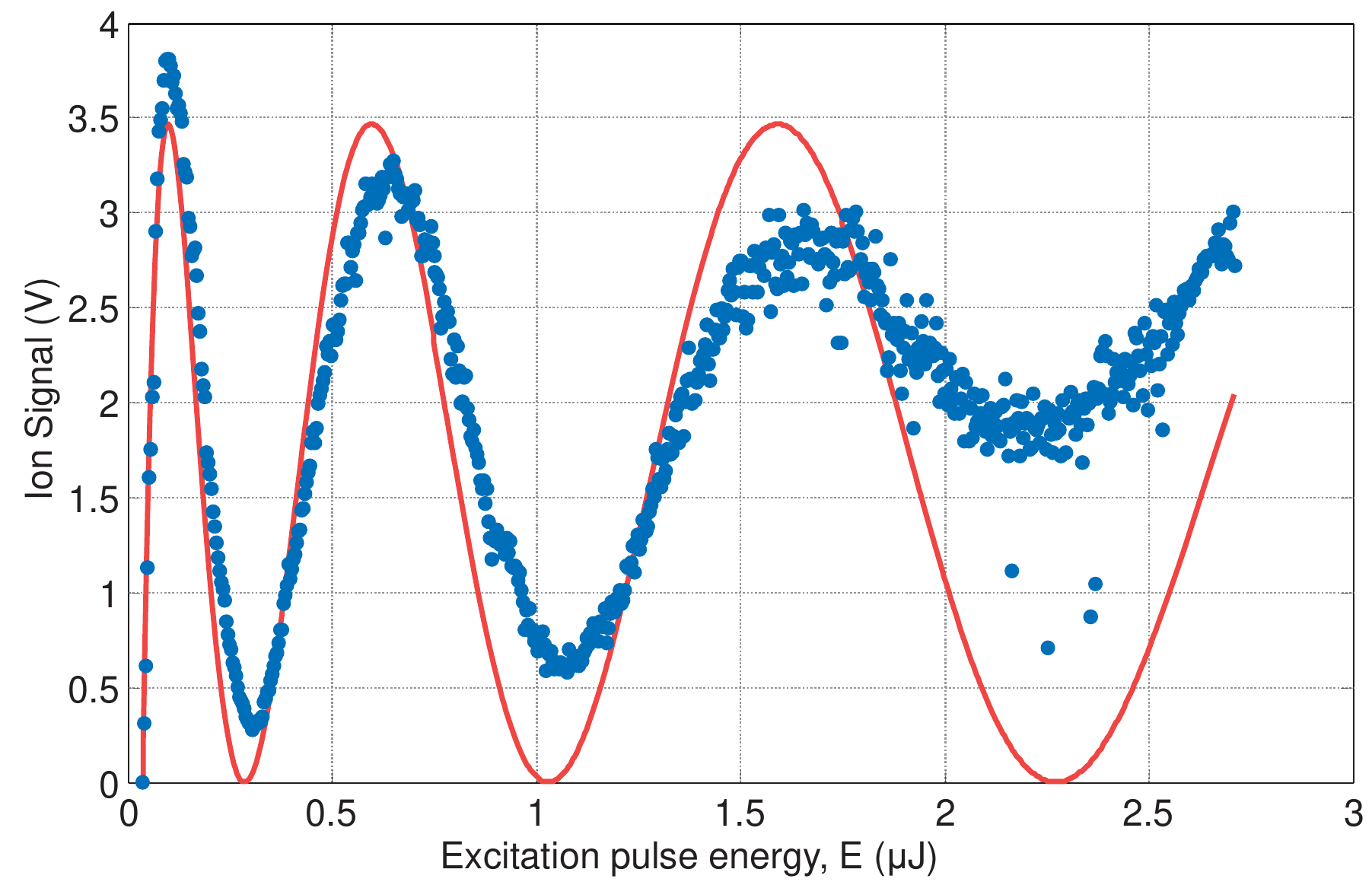}
\caption{(Color online). Rb$^{+}$ ion signal (blue dots), proportional to the population of $5p_{1/2}$ state of Rb, as a function of the excitation pulse energy $E$. Red solid line shows the best fit by $\sin^{2}(A/2)$, with $A$ being the pulse area calculated for a given pulse energy.}
  \vskip -.1truein
  \label{FigResultRbRabi}
\end{figure}

We first observed Rabi oscillations in a two-level system in order to test our experimental setup, and calibrate the pulse area. We used the D$_{1}$ transition ($5s_{1/2} \rightarrow 5p_{1/2}$) of atomic Rb at 794.75 nm. Initial bandwidth of our excitation pulses was narrowed from 10.3 nm to 6.2 nm full width at half maximum (\textsc{fwhm}). At this bandwidth the neighboring $5s_{1/2} \rightarrow 5p_{3/2}$ transition at 780 nm can be safely disregarded (as confirmed by our numerical analysis). Pulse energy was attenuated and scanned by a variable neutral density filter.  The probe wavelength was tuned to 1300 nm. The observed dependence of the ionization rate, and therefore population of the $5p_{1/2}$ state, on the excitation energy is shown in Fig.\ref{FigResultRbRabi}. As expected, the excited state population exhibits oscillatory behavior, representative of Rabi oscillations. The intensity of pump pulses at the first minimum of oscillations, calculated from the measured energy and beam diameter, was about $2.1\times 10^{9}$ W/cm$^2$. Given the transition dipole moment of 2.53 Cm \cite{NIST, LoudonBook}, this intensity corresponds to the pulse area of $2.2\pi $ (main uncertainty associated with the beam diameter value), in reasonable agreement with the experimental value of $2\pi $.

We have found that successful observation of femtosecond Rabi oscillations depends critically on the following conditions. First and most important, the pulse area of any unknown field preceding the excitation pulse (so-called ``pre-pulse'', often generated by regenerative amplifiers) must be much smaller than $\pi$. Even a relatively weak pre-pulse will transfer some population to the excited state, preparing atoms in a superposition state. In the event that the optical phase of such a pre-pulse differs from that of the main excitation pulse, the contrast of Rabi oscillations will degrade with increasing pulse energy. Though we managed to suppress the energy of a pre-pulse (found in our case 2.8 ns ahead of the main pulse) to less than 1\% of the main pulse energy, this proved satisfactory only for observing the first two periods of Rabi oscillations. We attribute the decay of the oscillation amplitude with increasing pulse energy (clearly seen in Fig.\ref{FigResultRbRabi}) to the presence of a residual pre-pulse.

The second critical point of concern is related to the spectral phase distortions of the excitation pulse which result in a time-dependent instantaneous frequency and hence affect the dynamics of Rabi oscillations. Though the lowest-order (quadratic with frequency) distortions can be eliminated by means of a pulse compressor, higher orders have to be compensated with an external pulse shaper. We carried out such compensation using the technique of multi-photon intra-pulse interference phase scans (MIIPS) \cite{Lozovoy04}, achieving the spectral phase flatness of better than 0.25 radian across \textsc{fwhm} of the excitation spectrum. Decreasing contrast of Rabi oscillations is partly attributed to the residual phase distortions.

Finally, the non-uniform spatial distribution of laser intensity in a focused Gaussian beam results in smearing out Rabi oscillations when the effect is averaged over the full beam profile. We minimized such averaging by tightly focusing our probe beam into the central part of the excitation region (ratio between the beam diameters of 0.4 and 0.3 for experiments with Rb and K, respectively) and therefore sampling the population in the region of relatively uniform intensity. In addition, multiphoton nature of ionization from the excited state effectively reduces the spatial area of probing.

To demonstrate Rabi oscillations between a single ground state and a coherent superposition of several excited states, we used the D lines of atomic potassium, $4s_{1/2} \rightarrow \{4p_{1/2}, 4p_{3/2}\}$, with transition wavelengths of 769.9 nm and 766.5 nm, respectively. The difference between these wavelengths was well within the bandwidth of our laser pulses, which allowed simultaneous excitation of both transitions. The central wavelength of the excitation pulse was tuned to 768.2 nm. The superposition of $4p_{1/2}$ and $4p_{3/2}$ was probed by the subsequent photo-ionization with a broadband probe pulse at 1254 nm. The probe central wavelength was tuned on resonance with an intermediate state $5s_{1/2}$ (Fig.\ref{FigSetup} (c)). The interference of two interaction paths, $4p_{1/2} \rightarrow 5s_{1/2}$ and $4p_{3/2} \rightarrow 5s_{1/2}$, results in the appearance of quantum beating in the ionization signal as a function of the time delay between the excitation and probe pulses, as discussed below. The dipole moments for the $4p_{1/2} \rightarrow 5s_{1/2}$ and $4p_{3/2} \rightarrow 5s_{1/2}$ transitions are different so the probe wavelength was varied in order to equalize the transition probabilities and therefore to maximize the contrast of quantum beatings.
\begin{figure}[t]
\centering
\includegraphics[width=0.98\columnwidth]{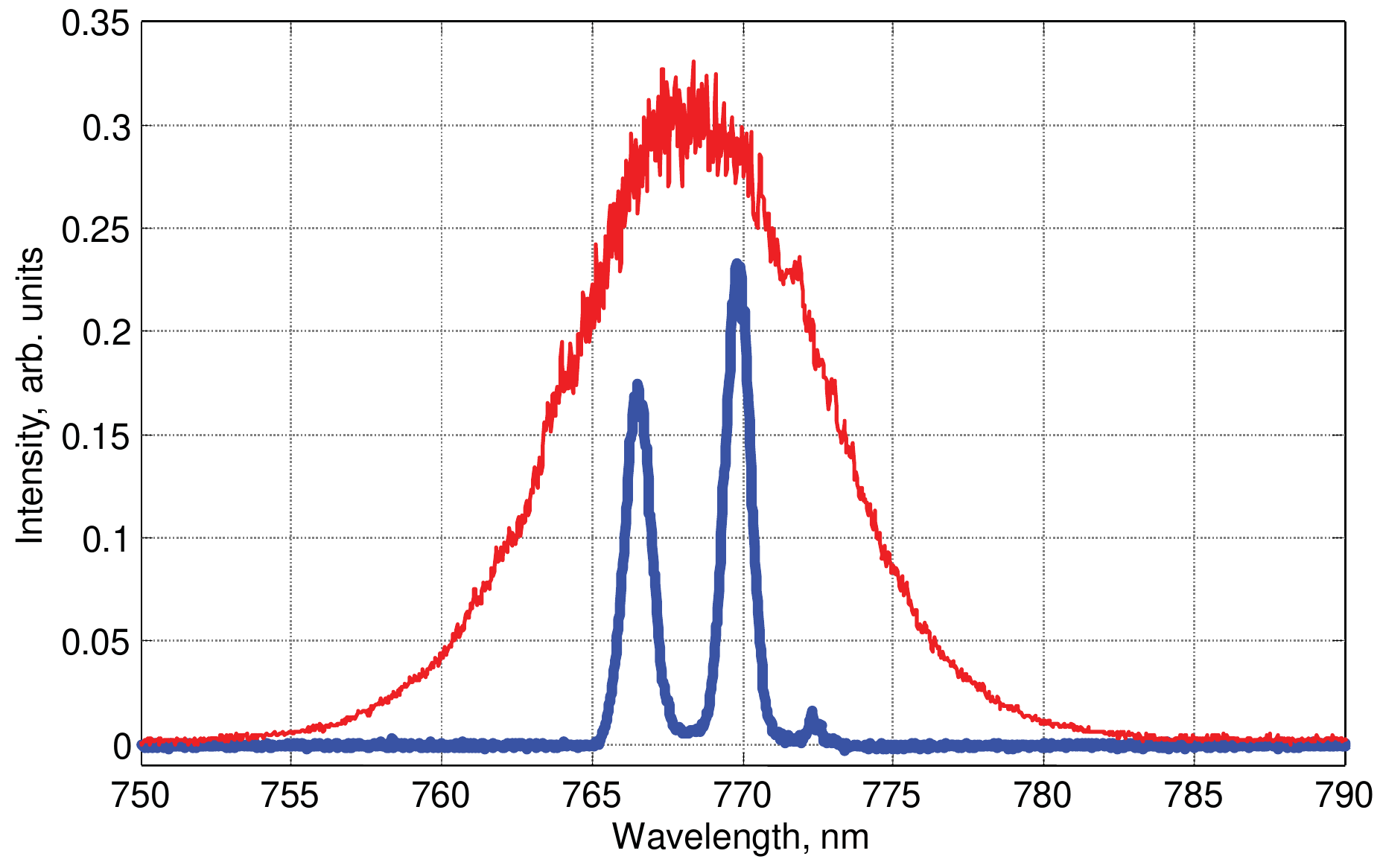}
\caption{(Color online). Pulse spectrum for exciting a coherent superposition of $4p_{1/2}$ and $4p_{3/2}$ states of Potassium (thick blue), obtained by means of spectral pulse shaping of the original broadband pulse (thin red). Two main peaks are centered at the corresponding resonant frequencies. A weak line at 772 nm is due to the imperfections of the pulse shaper. It has been included in our numerical analysis and resulted in less than $5\%$ change in the final state populations for the pulse energies used in the experiment.}
  \vskip -.1truein
  \label{FigSpectrum}
\end{figure}

Using the pulse shaper, we blocked all frequencies in the excitation spectrum except for two windows around 766.5 nm and 769.9 nm (Fig.\ref{FigSpectrum}). Each spectral window of 1.8 nm \textsc{fwhm} corresponded to a transform limited pulse of about 0.5 ps length. The contrast of the observed quantum beats was maximized by varying the amplitude ratio of the two resonant spectral peaks.
\begin{figure}[t]
\centering
\includegraphics[width=0.98\columnwidth]{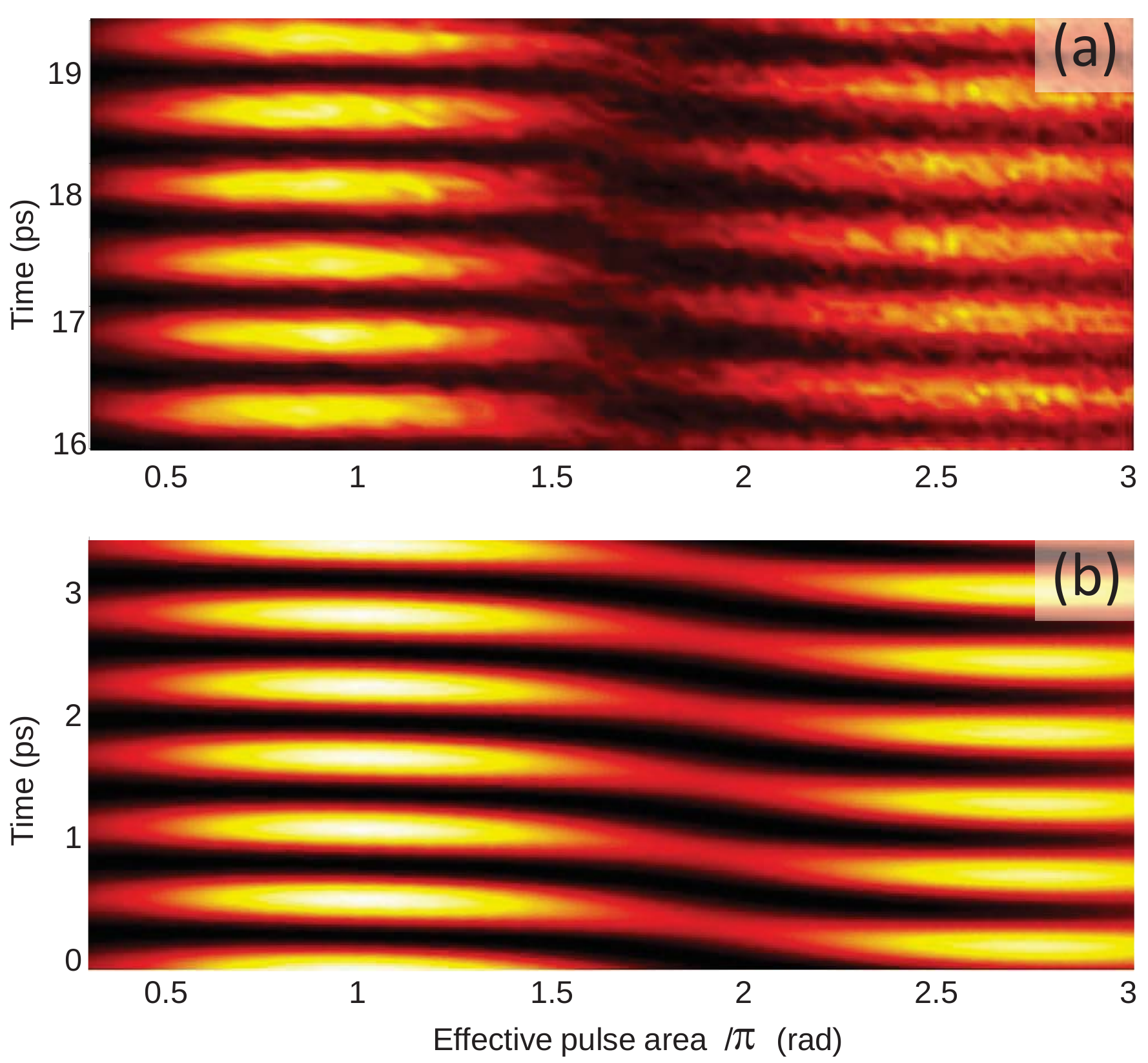}
\caption{(Color online). Experimental results (\textbf{a}) and numerical simulations (\textbf{b}) of Rabi oscillations between a single state and a wavepacket. The two-dimensional plots show the ion signal (color coded) as a function of the effective pulse area and probe delay.}
  \vskip -.1truein
  \label{FigNumerics}
\end{figure}

\begin{figure}[t]
\centering
\includegraphics[width=0.98\columnwidth]{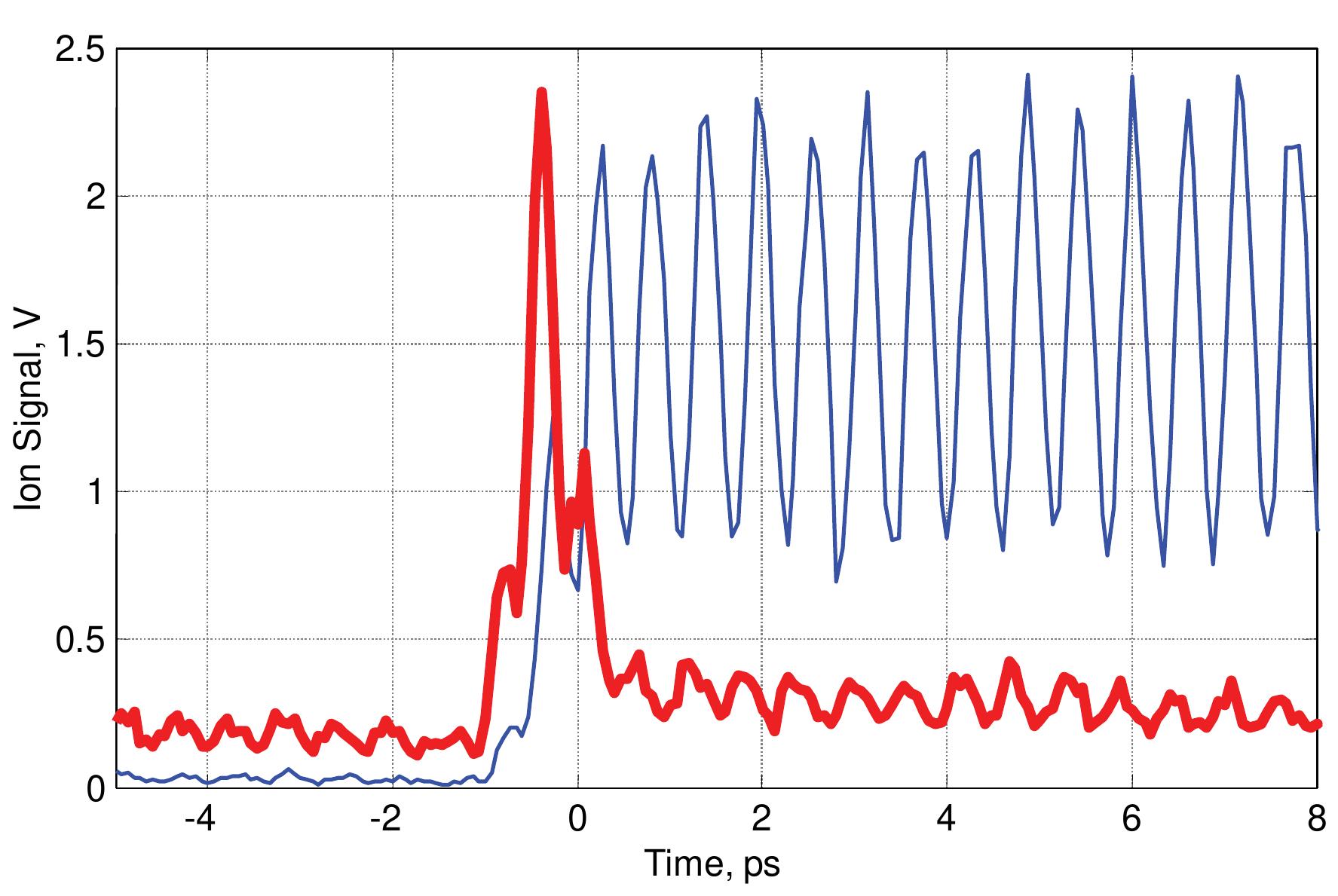}
\caption{(Color online). Quantum beating of the excited wavepacket for two excitation pulse areas, $\pi $ (thin blue) and $2\pi $ (thick red).}
  \vskip -.1truein
  \label{FigResultsBeat}
\end{figure}

The main result of this work is shown in Fig.\ref{FigNumerics} (a). Color coded ion signal is plotted as a function of the effective pulse area (horizontal axis) and time delay between the excitation and probe pulses (vertical axis). The effective pulse area is defined as $A_{\emph{eff}}=\sqrt{A_{1}^2+A_{2}^2}$ where $A_{1,2}$ are the pulse areas for the two individual transitions $4s_{1/2}\ \rightarrow 4p_{1/2}$ and $4s_{1/2} \rightarrow4p_{3/2}$, respectively (for reasons behind this definition of $A_{\emph{eff}}$, see \cite{Shapiro09}). As expected, the ionization signal exhibits oscillations along the vertical axis, indicative of quantum beats between the $4p_{1/2}$ and $4p_{3/2}$ states of Potassium. Repetitive appearance and disappearance of the quantum beat signal with changing excitation energy (horizontal axis) is the result of Rabi oscillations between the ground state and the coherent superposition of two excited states. This can be seen more clearly by examining the beating signal at two different excitation energies corresponding to pulse areas of about $\pi $ and $2\pi $, shown in Fig.\ref{FigResultsBeat} by thin blue and thick red curve, respectively. Relatively high contrast of the quantum beats attests to the efficient population transfer to both excited states. The Fourier spectrum of the beat signal shows a strong peak at 1.77 THz, in close agreement with the fine structure splitting of 1.73 THz. Non-zero signal at negative time delays, clearly seen in the case of a higher energy excitation ($2\pi $), is due to the laser pre-pulse discussed above.

We calculated the wavefunction amplitudes of $4s_{1/2}$, $4p_{1/2}$ and $4p_{3/2}$ by numerically solving the Schr\"{o}dinger equation (for details, see \cite{Zhdanovich09}). In Fig.\ref{FigNumerics}(b), the numerical results are plotted in a two-dimensional form equivalent to that used in panel (a). The effective pulse area is scanned along the horizontal axis, whereas the relative phase between the two excited states changes along the vertical axis, and their total population is color coded. Good agreement between the measured and calculated signals enabled us to utilize the latter for better understanding of Rabi oscillations which involve multiple excited states.

Two effects, observed in the experiment and confirmed by the numerical calculations, are seen in Fig.\ref{FigNumerics}(b). First, the ionization signal does not drop to zero after one full oscillation. Second, the phase of a wavepacket depends on the excitation energy. Both effects can not be explained by a simplified picture in which each excited state interacts only with the corresponding single resonant component of the driving field while being insensitive to the off-resonant part.

\begin{figure}[t]
\centering
\includegraphics[width=0.8\columnwidth]{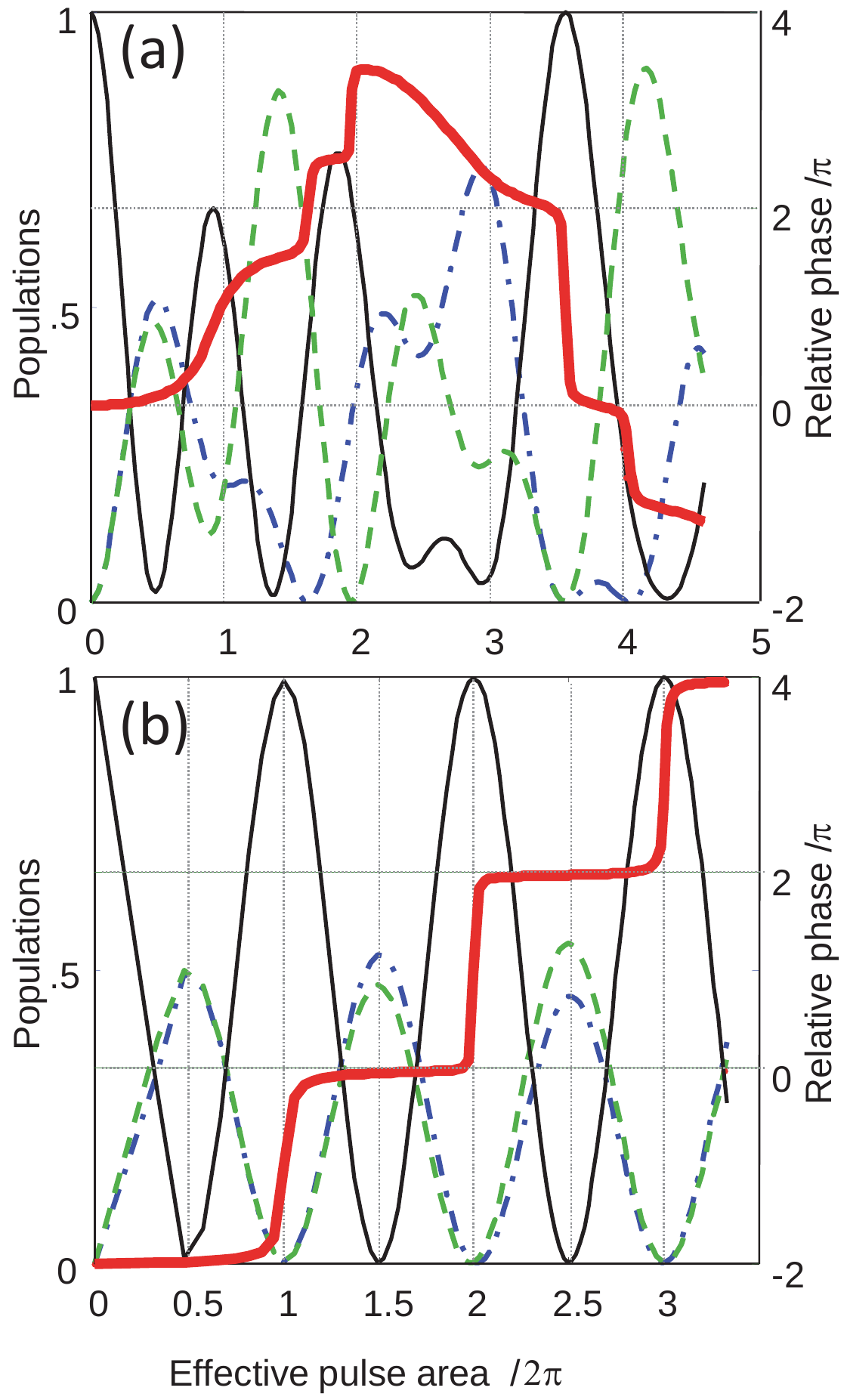}
\caption{(Color online). Numerical simulations of the interaction of Potassium with broadband laser pulses.  Similarly to our experimental conditions, the excitation spectrum consists of two separate resonant peaks (Fig.\ref{FigSpectrum}). The spectral width of each peak is 1.8 nm (a) and 0.36 nm (b). Populations of the two excited states, $4p_{1/2}$ (dash-dotted blue) and $4p_{3/2}$ (dashed green), and their relative phase (thick solid red) are plotted as a function of the effective pulse area of the excitation field. Ground state population is shown as a solid black line.}
  \vskip -.1truein
  \label{FigNumerics2}
\end{figure}

Fig.\ref{FigNumerics2} shows the calculated populations of the two quantum states, composing an excited wavepacket, together with their relative phase as a function of the effective pulse area. Panel (a) corresponds to the experimentally used excitation spectrum with two resonant peaks being 1.8 nm broad (Fig.\ref{FigSpectrum}). In panel (b), the spectral peaks were narrowed to 0.36 nm. Narrower bandwidth implies longer pulse and weaker electric field for a given pulse area. Since the Stark shift scales quadratically with the amplitude of the off-resonant electric field, narrowing bandwidth effectively decreases the shift of $4p_{1/2}$ caused by the spectral peak resonant with $4p_{3/2}$, and vice versa. Hence, the result of panel (b) can be easily explained by a simplified interaction picture in which off-resonant coupling is neglected \cite{Gasiorowicz}. Excited state populations oscillate in-phase and return to zero at $2\pi$ pulse area. In contrast, panel (a) shows more complex oscillatory dynamics, which result in an incomplete return of population to the ground state, as observed in our experiment. This strong-field effect may prove especially important in the case when the contrast of Rabi oscillations is utilized for assessing the degree of coherence of an atom-photon interaction.

The relative phase between the two excited states in a wavepacket (solid red line in Fig.\ref{FigNumerics2}) deserves a closer look, since many applications of Rabi oscillations, e.g. in quantum computing, rely heavily on this phase behavior. The latter can be described by the following qualitative picture. Consider a system of two levels, $|g\rangle$ and $|e\rangle$, interacting with a cw laser field. At any given time, the wavefunction of this system can be expressed as:
\begin{eqnarray}
\label{EqWaveFunction}
\Psi(t)&=& a(t)|e\rangle e^{-i(\omega_{0} t +\delta(t))}+b(t)|g\rangle,
\end{eqnarray}
where $a(t)$ and $b(t)$ are real amplitudes, $\delta(t)$ is the time dependent relative phase between the two quantum states, and $\omega_{0}$ is the transition frequency. By solving Maxwell-Bloch equations describing the dynamics of this system in the laser field \cite{AllenEberlyBook}, one finds for the excited state population $a^{2}(t)$ and relative phase $\delta(t)$:
\begin{eqnarray}
\label{EqPopulationPhase}
a^2(t)&=&\frac{\Omega_0^2}{\Omega^2}\sin^2\frac{\Omega t}{2} \cr
\delta(t)&=-&\arctan(-\frac{\Omega}{\Delta}\cot(\frac{\Omega t}{2}))-\Delta t,
\end{eqnarray}
where $\Omega_0 $ is the intensity dependent resonant Rabi frequency, $\Delta $ is the detuning of the excitation field frequency from $\omega_{0}$, and $\Omega=\sqrt{\Omega_0^2+\Delta^2}$.
One can see from Eq. \ref{EqPopulationPhase} that every time the excited state population passes through zero, the relative phase between the two states undergoes a jump by $\pi $ radian. From our numerical analysis, we conclude that similar behavior holds for a system of multiple (in our case, two) excited states. Namely, when the population of any of the excited states reaches zero, the phase of the corresponding wavefunction and therefore the phase of a wavepacket exhibits a ``$\pi$-jump'' (e.g. at pulse areas of $1.7\pi$ and $1.9\pi$ in Fig.\ref{FigNumerics2}). When two zero crossings coincide (here, at approximate pulse area of $3.5\pi$) the phase jumps add up resulting in a larger overall change of the wavepacket phase.

In summary, we have experimentally demonstrated Rabi oscillations of atomic population between a single ground electronic state and a coherent superposition of two excited states, executed with broadband ultrashort pulses. Our results confirm the feasibility of applying $\pi $-pulses to multi-level systems for efficient population transfer on a femtosecond time scale. We illustrate, both experimentally and numerically, the limitations of ultrafast population transfer due to the Stark shifts resulting from strong off-resonant interactions. We analyze the phase of an excited wavepacket and present a simple picture explaining its complex dependence on the excitation pulse area.

\begin{acknowledgements}
This work has been supported by the CFI, BCKDF and NSERC.
\end{acknowledgements}


\end{document}